\documentstyle[aps,12pt]{revtex}
\begin{document}
\title{Statistical analysis and modeling of variations of the earth's magnetic field}
\setcounter{page}{1} 
\author{Jon D. Pelletier}
\address{
  Department of Geological Sciences, Snee Hall, Cornell University \\
  Ithaca, NY 14853 \\
}
\maketitle 

\begin{abstract}
Power spectral analyses of the dipole moment of the earth's magnetic field
inferred from 
ocean sediment cores and archeomagnetic data from
time scales of 100 yr to 4 Myr have been carried out. The power spectrum
is proportional to $1/f$ where $f$ is the frequency. 
These analyses compliment
previous work which has established a 
$1/f^{2}$ spectrum for variations at time
scales less than 100 yr. Power spectral analyses of inclination and
declination inferred from lake sediments from time scales of 10 yr to
30 kyr have also been performed. The spectra are constant above time scales
of 3 kyr, proportional to $1/f^{2}$ from time scales of 500 yr to 3 kyr,
and constant again below time scales of 500 yr. The 3 kyr time scale is
associated with the decay time of the quadrupole moment. We test the
hypothesis that reversals are the result of variations in dipole intensity
with a $1/f$ spectrum which occasionally are large enough to cross the
zero intensity value. Synthetic binormal
time series with a $1/f$ power spectrum representing variations in the 
earth's dipole moment are constructed. 
Synthetic reversals from these time series exhibit statistics
in good agreement with the reversal record. 
$1/f$ noise behavior is
reproduced with a model of magnetic diffusion in the earth's core 
driven by dynamo action modeled as a random amplification or destruction of the
local magnetic field. 
\end{abstract}

\newpage
\section{Introduction}

The earth's magnetic field has exhibited significant variability over a wide range
of time scales. On time scales less than a couple of hundred years
historical data are available for variations in the intensity and orientation
of the geomagnetic field. Archeomagnetic data can be used to infer the 
intensity of the field from time scales of centuries to millenia. Sediment
cores provide the widest range of time scales of variations in the
geomagnetic field with internal origin: 1 kyr to several Myr. Techniques of time
series analysis can be used to characterize this variability. The power spectrum
is the square of the coefficients of the Fourier transform of the time
series. It quantifies the average variability of the series at different 
time scales. 
Barton (1982) 
has performed spectral analysis of paleointensity using historical observations
and sediment cores. He identified a broad, continuous power spectrum 
with a steep dependence on time scale for time scales less than a century and
a flatter spectrum at longer time scales.     

The geomagnetic field also exhibits reversals with a complex history including
variations over a wide range of time scales. The reversal history can be 
characterized by the polarity interval distribution and the reversal rate.
Polarity intervals vary from those short enough to be barely resolved in
the magnetic anomalies of the seafloor to the 35 Myr Cretaceous superchron. 
Reversals are also clustered in time such that short polarity intervals
tend to be followed by short polarity intervals and long intervals
by long intervals. This clustering has been quantified with the reversal rate which
gradually decreases going back to 100 Ma and then increases going back further in
time before the 
Cretaceous superchron. 

Due to the availability of many new time series data sets for
paleointensity and inclination and declination of the earth's magnetic field
since the work of Barton (1982), 
it would be useful to perform power spectral analyses of some
of these recent data sets to further characterize the temporal variability of
the geomagnetic field.
In this paper we perform power spectral analyses of time series data for
the dipole moment and the inclination and declination of the earth's magnetic 
field inferred from sediment core and archeomagnetic data. We find that
the power spectrum of the virtual axial dipole moment (VADM)
from time scales of 100 yr to 4 Myr is well approximated
by a $1/f$ dependence, where $f$ is the frequency. 
The power spectrum
of inclination and declination are constant above time scales
of 3 kyr, proportional to $1/f^{2}$ from time scales of 500 yr to 3 kyr,
and constant again below time scales of 500 yr.
Variations in the intensity of the geomagnetic field in
one polarity exhibits a normal distribution. When a fluctuation crosses the
zero intensity value a reversal occurs. We test the hypothesis that reversals
are the result of intensity variations with a $1/f$ power spectrum 
which occasionally are large enough
to cross the zero intensity value, driving the geodynamo into the opposite 
polarity state. Synthetic time series with a
$1/f$ power spectrum and a binormal distribution
are used to generate reversal statistics. These
are found to be in good agreement with those of the real reversal history.
The reversal statistics
are sensitive to the form of the power spectrum of intensity variations. 
We conclude that the
agreement between the synthetic reversal record produced with $1/f$ noise
variations in the earth's dipole moment
is strong support for $1/f$ behavior over the length of the reversal
record. This suggests that processes internal to the core may determine geomagnetic
variability up to very long time scales. This contrasts with the hypothesis
that internal core processes
dominate secular variations while changes in conditions at the core-mantle 
boundary determine variations at larger time scales
(McFadden and Merrill, 1995). 
A model of geomagnetic variability based on internal processes of
magnetic diffusion and dynamo action modeled as a stochastic process
generates the observed
$1/f$ behavior. This model is analyzed in detail
and compared to the behavior of the earth's magnetic field.

\section{Power Spectral Analyses of the Virtual Axial Dipole Moment}
Paleomagnetic studies clearly show that the polarity
of the magnetic field has been subject to reversals.
Kono (1971) 
has compiled paleointensity measurements of the magnetic field from volcanic
lavas for 0-10 Ma.
He concluded that the distribution of
paleointensity is well approximated by a symmetric binormal
distribution
with mean 8.9x10$^{22}$Am$^{2}$ and standard deviation 3.4x10$^{22}$Am$^{2}$.
One normal distribution is applicable to the field when it is in
its normal polarity and the other when it is in its reversed polarity.
 
We have utilized three datasets for computing the power spectrum of
the dipole moment of the earth's magnetic field. 
They are archeomagnetic data from time scales
of 100 yr to 8 kyr from Kovacheva (1980), marine sediment
data from the Somali basin from time scales of 1 kyr to 140 kyr from
Meynadier et al. (1992), and marine sediment data from the Pacific and
Indian Oceans from 20 kyr to 4 Myr from Meynadier et al. (1994).
The data were published in table form in
Kovacheva (1980) and obtained from L. Meynadier (Meynadier, 1995)
for the marine sediment
data in Meynadier et al. (1992) and Meynadier et al. (1994).
Marine sediment data are accurate measures of
relative paleointensity but give no information on absolute intensity.
In order to calibrate marine sediment data, the data must be compared to absolute
paleointensity measurements from volcanic lavas sampled from the same time
period as the sediment record. Meynadier et al. (1994) has done this
for the composite Pacific and Indian Ocean dataset. They have calibrated the
mean paleointensity in terms of the virtual axial dipole moment
for 0-4 Ma as 9x10$^{22}$Am$^{2}$
(Valet and Meynadier, 1993). This value is
consistent with that obtained by Kono (1971) for the longer time
interval up to 10 Ma. Using this
calibration, we calibrated the Somali data with the time interval 0-140 ka from the
composite Pacific and Indian Ocean dataset.
The data from Meynadier et al. (1994) are plotted in Figure 1 as a function of
age in Ma. The last reversal at approximately 730 ka is clearly shown.
We computed the power spectrum of each of the time series with the Lomb
periodogram (Press et al.,
1992). 
The compiled spectra are given in Figure 2. 
The composite sediment record from the Pacific and
Indian Oceans are plotted up to a frequency corresponding to a period
of 25 kyr. Above this time scale good synchroneity is observed
in the Pacific and Indian Ocean datasets (Meynadier et al., 1994). This
suggests
that non-geomagnetic effects such as variable sedimentation rate
are not significant in these cores above this time scale.
From frequencies corresponding to time scales of 25 kyr down to 1.6 kyr
we plot the power spectrum of the Somali data.
From time scales of 1.6 kyr to
the highest frequency we plot the power spectrum of the
data of Kovacheva (1980). 
A least-squares linear regression to the data yields a
slope of $-1.09$ over $4.5$ orders of magnitude. This indicates that
the power spectrum is well approximated as $1/f$ on these time scales.

The power spectrum of secular geomagnetic intensity variations has been
determined to have a $1/f^{2}$ power spectrum between time scales of one
and one hundred years (Currie, 1968; Barton, 1982;
Courtillot and Le Mouel, 1988). This is consistent with the
analysis of McLeod (1992) who found that the first difference of annual
means of geomagnetic field intensity is a white noise since
the first differences of a random process with power spectrum $1/f^{2}$ 
is a white noise. 
Our observation of $1/f$ power spectral behavior above time scales
of approximately 100 yr together with the results of Currie (1968) and
Barton (1982)  
suggests that there is a crossover from $1/f$ to $1/f^{2}$ spectral
behavior at a time scale of approximately
one hundred years.
 
\section{Analysis of the Reversal History}
In this section we will test the hypothesis that reversals are the 
result of intensity variations in one polarity state becoming large enough
to cross the zero intensity value into the opposite polarity state. 
We will 
show that the statistics of the reversal record are 
consistent those of a binormal, $1/f$ noise paleointensity record which 
reverses when the intensity crosses the zero value.
We will compare the polarity length distribution
and the clustering of reversals between synthetic reversals produced with
$1/f$ noise intensity variations and the
reversal history according to Harland et al. (1990) and Cande and Kent 
(1992a,1995).
 
First we consider the polarity length distribution of the real reversal 
history. The polarity length distribution calculated from the chronology of
Harland et al. (1990) is given as the solid line in Figure 3. The polarity
length distribution is 
the number of interval lengths
longer than the length plotted on the horizontal axis. 
A reassessment of the magnetic anomaly data
has been performed by Cande and Kent (1992a,1995) to obtain an alternative
magnetic
time scale. Their chronology, normalized to the same length as the Harland
et al. (1990) time scale, is presented as the dashed curve. The two distributions
are nearly identical.
These plots suggest that the polarity length distribution is better fit
by a power law for large polarity lengths than by an exponential distribution,
as first suggested by Cox (1968).
The same conclusion has been reached by Gaffin (1989) and Seki and Ito
(1993). It should be emphasized, however, that the polarity length distribution
is a very different analysis than that of the spectral analysis of Section 
2. There is no simple relationship between the power spectrum of intensity
variations and the polarity length distribution of reversals. 
The observation of a power-law power spectrum in Section 2 does not
necessarily imply a power-law polarity length distribution. 
 
The third curve, plotted with a dashed and dotted line, represents
the polarity length distribution estimated from the magnetic time scale between
C1 and C13 with ``cryptochrons'' included and scaled to the length of the
Harland et al. (1990) time scale. Cryptochrons are small variations
recorded in the magnetic anomaly data that may either represent variations in
paleomagnetic intensity or short reversals (Blakely, 1974;
Cande and Kent, 1992b).
Cryptochons occur with a time scale at 
the limit of temporal resolution of the reversal record from 
magnetic anomalies of the sea floor.
The form of the polarity length distribution estimated from the record between
C1 and C13
including cryptochrons is not representative of the entire reversal history
because of 
the variable reversal rate which
concentrates many short polarity intervals in this time period. However, this
distribution enables us to estimate the temporal resolution of the reversal
record history. The distribution estimated from C1 to C13
has many more short polarity intervals than those of the full reversal
history starting at a reversal length of 0.3 Myr. Above a time
scale of 0.3 Myr the magnetic time scale is nearly complete. Below it many
short polarity intervals may be unrecorded.
 
To show that the polarity length distribution of the real reversal record is 
consistent with that produced by binormal,
$1/f$ noise variations in the earth's dipole moment, we have generated
synthetic Gaussian noises with a power spectrum porportional to $1/f$,
a mean value of 8.9x10$^{22}$Am$^{2}$
and a standard deviation of 3.4x10$^{22}$Am$^{2}$.
These synthetic noises represent the field intensity in one
polarity state. The synthetic noises were generated using
the Fourier-domain filtering technique described in Turcotte (1992).
An example is shown in Figure 4a.
In order to construct a binormal intensity distribution from the synthetic
normal distribution, we inverted every other polarity interval to the
opposite polarity starting from its minimum value below the zero intensity
axis and extending to its next minimum below the zero. The result of this
procedure on the Gaussian, $1/f$ noise of Figure 4a is presented
in Figure
4b.
Its irregular polarity lengths are similar to those in the marine sediment
data of Figure 1.
 
The operation of reversing the paleomagnetic intensity when it crosses the
zero intensity value is consistent with models of the geodynamo as a
system with two symmetric attracting states of positive and negative
polarity such as the Rikitake disk dynamo (Rikitake, 1958).
Between reversals, the
geomagnetic field fluctuates until a fluctuation large enough occurs
to cross the energy barrier into the other basin of attraction. Kono (1987)
has explored the statistical similarity between the Rikitake disk dynamo
and the distribution of paleointensity. Our
construction of the binormal $1/f$ noise is consistent with this model.
 
We have
computed the distribution of lengths between successive reversals
for twenty synthetic noises scaled to 170 Ma, the length of the reversal
chronology, and averaged the
results in terms of the number of reversals. The results are given in Figure
5. The dots
are the maximum and minimum values
obtained in the twenty synthetic reversal
chronologies, thus
representing 95\% confidence intervals. The polarity interval
distribution from the Harland et al. (1990) time scale indicated
by the dashed curve falls
within the 95\% confidence intervals of our synthetic data
over all time scales plotted except for the Cretaceous
superchron, which lies slighly outside of the 95\% confidence interval
and reversals
separated by less than about 0.3 Myr.
The overprediction of very short
reversals could be a limitation of the model or a result of the incompleteness
of the reversal record for short polarity intervals.
As mentioned, the temporal resolution of the magnetic time scale inferred from
magnetic anomalies is approximately 0.3 Myr. We conclude that the 
polarity length distribution produced from binormal $1/f$ intensity variations
are consistent with the observed 
polarity length distribution for all time scales at which the reversal record
is complete.
 
We next consider whether the agreement illustrated
in Figure 5 is unique to $1/f$ noise. We have computed polarity
length distributions using the binormal intensity variations
with power spectra $f^{-0.8}$ and $f^{-1.2}$. These results along with
the $1/f$ result from Figure 5 are given in Figure 6. 
The shape of the polarity length distribution is very sensitive to the
exponent of the power spectrum. A slight increase in the magnitude of
the exponent results in many more long polarity intervals than with
$1/f$ noise. We conclude that the agreement in Figure 5 between the 
synthetic reversal distribution and the true reversal history is unique
to $1/f$ noise and provides strong evidence that the dipole moment has
$1/f$ behavior up to 170 Ma. 
 
A binormal, $1/f$ noise geomagnetic field variation is consistent with the
qualitative results of Pal and Roberts (1988) who found an
anticorrelation
between reversal frequency and paleointensity. This anticorrelation is
evident in the synthetic $1/f$ noise of Figure 4b. During the time
intervals
of greatest average paleointensity the reversal rate is lowest.
 
In addition to the broad distribution of polarity lengths, the reversal 
history is also characterized by a clustering of reversals. This behavior
has been quantified with the reversal rate. The reversal rate has 
been relatively high
from 0-20 Ma and has decreased gradually going back in history
to the Cretaceous superchron. 
An alternative approach to quantifying the clustering of reversals is
with the pair correlation function. 
The pair correlation function $c(t)$ is the
number of pairs of reversals whose separation is between $t$ and $t+\Delta t$,
per unit time (Vicsek, 1992). 
The pair correlation function for a 
set of points can be compared to that for a Poisson process to detect
non-random clustering. 
The pair-correlation function analysis is more 
appropriate for comparison of the reversal history to the synthetic reversal
history generated by a stochastic model such as model based on  
turbulent processes in the core. 
This is because stochastic models cannot predict behavior in time, such as
when the reversal rate is large or small.
However, a stochastic model 
may accurately reflect the extent to which small polarity intervals
are followed by small polarity intervals and long intervals by long intervals
as quantified with the pair correlation function.

The pair correlation function of
reversals according to the Harland et al. (1990) and Cande and Kent (1992a,1995)
reversal history
are shown in
Figure 7 as filled and unfilled circles, respectively. 
Also presented in 
Figure 7 is the pair correlation function for a synthetic
reversal data set based on binormal $1/f$ noise dipole moment variations
(boxes) and the pair correlation function for a Poisson process (triangles). 
The functions are offset so that they may be placed on the same graph.
The Poisson process was constructed with 293 points, the same number of reversals
as the Harland et al. (1990) time scale,
positioned with uniform probability on the interval between
0 and 170 Ma. The Poisson process yields a correlation function
independent of separation.
The real and synthetic reversal histories 
variations exhibit significant clustering with more pairs of points at
small separation and fewer at large separations than for a Poisson process. 
Straight-line fits of the form $c(t)\propto t^{\alpha}$ were obtained. 
The purpose of this was to show that 
similar clustering is observed in the real and synthetic reversals. The exponents
$\alpha$
of the Harland et al. (1990), Cande and Kent (1992a,1995),
and synthetic reversals are -0.39, -0.31, and -0.42, respectively. 
Similar non-random clustering is observed in the real and synthetic reversals.
We conclude from the consistency between the polarity interval distribution
and pair correlation function of real reversals and those generated by
fluctuations of a bionormal $1/f$ noise process that there is strong evidence
for $1/f$ intensity variations for time scales up to the length of the reversal
record. 

\section{Power Spectral Analyses of Inclination and Declination}
Power spectral analyses of inclination and declination data have also been
carried out. We obtained time series data of inclination and declination
from lake sediment cores in the Global Paleomagnetic Database (Lock
and McElhinney, 1992). The core with the greatest number of data points
was from Lac du Bouchet (Thouveny et al., 1990). The power spectrum
of the inclination and declination at Lac du Bouchet
estimated with the Lomb Periodogram is
presented in Figure 8. We associate the spectra with a constant spectrum
below a frequency of $f\approx$ 1/(3 kyr) and a constant spectrum
above a frequency of $f\approx$
1/(500 yr). From frequencies of $f\approx$ 1/(3 kyr) to
$f\approx$ 1/(500 yr) the inclination and declination are Brownian walks
with $S(f)\propto f^{-2}$. Spectral analyses of inclination data from
five other sediment cores were calculated. These spectra are presented in
Figure 9. The spectra correspond, from top to bottom, to cores from
Anderson Pond (Lund and Banerjee, 1985), Bessette Creek (Turner et al.,
1982), Fish Lake (Verosub et al., 1986),
Lake Bullenmerri (Turner and Thompson, 1981),
and Lake Keilambete (Barton and McElhinny, 1985). Since the data sets have
fewer points there is more uncertainty in the spectra and they are characterized
by greater variability between adjacent frequencies. The spectra have the
same form, within the uncertainty of the spectra, as that associated with the
spectra from Lac du Bouchet. These results suggest that 3 kyr and 500 yr are 
characteristic time scales of geodynamo behavior.
Variations in inclination and declination are associated with changes in
the non-dipole components of the field.
Therefore, the autocorrelation or decay time of the quadrupole moment is the
maximum time scale for correlated fluctuations of inclination and declination
to occur. The autocorrelation time of the quadrupole moment has been estimated
by McLeod (1996) to be 1.6 kyr. This is within a factor of two
of the 3 kyr time scale above which variations in inclination and declination
are observed to be uncorrelated in the spectra of Figures 8 and 9.
 
Many analyses of variations 
in paleointensity of the earth's magnetic field concentrate on identifying
characteristic time scales of variation. Many such characteristic
time scales have been identified. Valet and Meynadier (1993) suggested,
based on the same sediment core data analyzed in this paper, that the
earth's magnetic field regenerates following a reversal on a time scale of
a few thousand years and then decays slowly on a time scale of 0.5 Ma
before the next reversal. They termed this an ``asymmetric saw-tooth'' pattern.
More recent data have shown that the ``asymmetric saw-tooth'' is not 
a robust pattern. 
Longer cores show a slow decay
preceding a reversal to be rare (Tauxe and Hartl, 1997). Moreover, Laj et al.
(1996) has shown that the magnetic field does not always regenerate quickly
after a reversal.
Thibal et al. (1995) have quantified the rate of decrease in field intensity
preceeding a reversal and found it to be inversely proportional to the length
of the polarity interval. The authors concluded from this
that the length of the reversal
was predetermined. Such behavior is not indicative of a predetermined polarity
length. This can be concluded by considering the null hypothesis that variations
in the field are characterized by any stationary random process.
By definition, a stationary time series has a variance which is independent
of the length of the series.
The average rate of change of the time series over a time interval
will then be a constant value divided by the interval of time, i.e. inversely
proportional to time interval. Therefore, any stationary random function satisfies
the relationship that Thibal et al. (1995) observed.

In the power spectral analyses
of geomagnetic variations inferred from sediment cores by Lund et al. (1988),
Meynadier et al.
(1992), Lehman et al.
(1996), and Tauxe and Hartl (1997)
dominant periodicities
in the record were identified and proposed as characteristic time scales
of geodynamo behavior.
However, it must be emphasized that any finite length record will
exhibit peaks in its power spectrum, even if the underlying
process is random, such as a $1/f$ noise. Periodicity tests such
as those developed by Lees and Park (1995) need to be applied to data
in order to assess the probability that a peak in a spectrum is statistically
significant. The periodicity tests developed by Lees and Park (1995)
are especially valuable because they do not depend on a particular model of the 
stochastic portion of the spectrum. 
Some of the periodicity tests that have been used in 
the geomagnetism literature assume forms for the stochastic portion of the 
spectrum that are not compatible with the $1/f$ process we have identified. 
See Mann and Lees (1996) for an application
of these techniques to climatic time series.
 
It is generally believed that secular geomagnetic variations are the result of
internal dynamics 
while longer time scale phenomena such as variations in the reversal
rate are controlled by variations in boundary conditions at
the core-mantle boundary (CMB) (McFadden and Merrill, 1995).
However, our
observation of continuous $1/f$ spectral behavior from time scales of 100 yr
to 170 Myr suggests that a single process controls variations in geomagnetic
intensity over this range of time scales. In Section 6 we consider a
model for geodynamo behavior which reproduces the $1/f$ dipole moment 
variations over a wide range 
of time scales and
exhibits many of the other features of geomagnetic variability
we have identified.
 
\section{Multipole Expansion of the Present Day Geomagnetic Field}
In addition to temporal variations of the field, detailed information on the
spatial structure of the present-day magnetic field from spherical harmonic
degree $n=1$ to $n=13$ is available. Above $n=13$ the magnetic field at the
earth's surface is dominated by crustal magnetic fields. 
Stevenson (1983) and Voorhies and Conrad (1996) have
presented the results of a spherical harmonic
expansion of the geomagnetic field extrapolated to the core-mantle boundary.
The results indicate that a broad spectrum of multipoles is
represented with an intensity decreasing with increasing order in the
spherical harmonic expansion. The spectrum is consistent with a
power-law function of wave number at the core-mantle boundary  
with exponent $-1$: $R_{n}\propto n^{-1}$. 
The form of the spectrum is not well constrained since an exponential fit, first
proposed by Langel and Estes (1982), and a power-law fit both match the observed 
spectrum equally well from $n=1$ to $n=13$. This is illustrated in Figure 1 of
McLeod (1996) where the two fits are compared.
Although the data
does not allow an unambiguous determination of the functional form of the
spectrum, Stevenson (1983) has shown that a spectrum
$R_{n}\propto n^{-1}$ is consistent with a multipole expansion expected based
upon other considerations.
Stevenson argues that an exponent close to $-1$ is the only one consistent
with energy conservation given the typical convective velocities and
the magnetic Reynolds number expected in terrestrial dynamos.
Voorhies and Conrad (1996) have argued that a $R_{n}\propto n^{-1}$
amplitude spectrum at the core-mantle boundary
does a superior job at predicting the location of the
core-mantle boundary and provides a better extrapolation to the observed
dipole field. In the next section we discuss a model of the earth's
magnetic field which
generates a field with 
a power spectrum $R_{n}\propto n^{-1}$. 

\section{Modeling of Variations in the Earth's Magnetic Field by Dynamo Processes}  
There has been great interest in $1/f$ noise processes in the physics
literature for many years (Weissman, 1988). 
One model of $1/f$ noise is a
stochastic
process comprised of a superposition of modes with exponential decay characterized
by different 
time constants. The time constant for a stochastic
process is defined through its autocorrelation function $a(\tau)$.
For a stochastic process with a single time constant $\tau_{o}$
the autocorrelation function is given by $a(\tau)=e^{-\frac{\tau}{\tau_{o}}}$.
The power spectrum of such a process is, by the Weiner-Khinchine theorem, the
Fourier transform of the autocorrelation function:
\begin{equation}
S(f)\propto\frac{\tau_{o}}{1+(2\pi f)^{2}}
\label{lorentzian}
\end{equation}
This is a Lorentzian spectrum with Brownian walk behavior ($S(f)\propto f^{-2}$) 
for time scales small
compared to $\tau_{o}$ and white noise
behavior ($S(f)=$ constant) above the characteristic time constant. 
If the stochastic process is composed of a superposition of modes with
time constants following a distribution
$D(\tau_{o})\propto\tau_{o}^{-1}$, where the $D(\tau_{o})\Delta\tau_{o}$ is the
net variance contributed by modes between $\tau_{o}$ and $\tau_{o}+\Delta\tau_{o}$,
then a $1/f$ spectrum results over a range of frequencies
(van der Ziel, 1950; Weissman, 1988).
Such a distribution of exponential
time constants has
been documented for the earth's magnetic field by McLeod (1996).
 
McLeod (1996)
calculated the autocorrelation of each degree of the geomagnetic field
during the last eighty years. The autocorrelation functions that he computed had
an exponential dependence on time with degree-dependent time constants
$\tau_{o}\propto n^{-2}$. 
This behavior 
is consistent with a diffusion process.
McLeod (1996)
attributed this autocorrelation structure
to a simple model of
the geomagnetic field in which the field was stochastically
generated with a balance between field regeneration and diffusive decay by
decay across a magnetic boundary layer. 
One way to model such a stochastic diffusion process
is with a two-dimensional diffusion equation driven by random noise: 
\begin{equation}
\frac{\partial B_{z}}{\partial t}=D\nabla^{2} B_{z} +\eta(x,y,t)
\end{equation}
where $B_{z}$ is the axial component of the magnetic field at a point inside the
core,
$\eta(x,y,t)$ is a Gaussian white noise representing random amplification
and destruction of the field locally by dynamo action. 
To this equation we add a term equal to $p-B_{z,tot}$:
\begin{equation} 
\frac{\partial B_{z}}{\partial t}=D\nabla^{2} B_{z} +\eta(x,y,t)+c(p-B_{z,tot})
\label{uh}
\end{equation}
where $c$ is a constant, $B_{z,tot}$
is the dipole moment integrated over all space, and 
$p$ is +1 if the dipole moment of the field outside the core-mantle boundary 
is positive and -1 if the dipole moment outside the core-mantle boundary is 
negative. 
The effect of this term is
to create two basins of attraction (polarity states) within which the dipole
field fluctuates around an intensity of +1 or -1 
until a fluctuation large enough occurs to cross the barrier
to the other basin of attraction. This term 
could be the result of a conservation of magnetic energy for the combination 
of the poloidal and toroidal fields
such that when the poloidal dipole field intensity 
is low the toroidal field 
intensity, which is unobservable outside the core and not explicitly modeled in 
equation \ref{uh}, 
is high and dynamo action is intensified, repelling the poloidal
field away from a state of low dipole intensity. 

In our model the core is
modeled as a two-dimensional circular region of 
uniform diffusivity (the fluid outer core) 
surrounded by
an infinite region with small but finite 
diffusivity and the boundary condition that $B_{z}$ approach zero as
$r$, the radial distance from the center of the earth, approaches infinity. 
The diameter of the inner circular region is the diameter of the 
core-mantle boundary. 
 
This model has been studied in terms of the distribution of values and power spectrum
of the dipole moment 
and the
power spectrum of the angular deviation from the dipole field.
The dipole field from the simulation is
plotted in Figure 10. The field clearly undergoes reversals with a broad
distribution of polarity interval lengths. Figure 11 presents the
dipole distribution of 10 simulations (solid curve) along with the fit to a binormal
distribution (dashed curve). A binormal distribution fits the distribution well.
The slight asymmetry is the result of this particular model run spending
slightly more time in the negative polarity state than the positive polarity
state. Model outputs were generated which showed asymmetry in the other direction. 
 
The average power spectrum of time series of the dipole field from 25
simulations is presented in Figure 12. The spectrum has a low-frequency
spectrum $S(f)\propto f^{-1}$ and a high-frequency spectrum $S(f)\propto f^{-2}$.
This is identical to the spectrum observed in sediment cores and historical
data discussed in Section 2. The crossover time scale is the diffusion time
across the diameter of the core, estimated to be between
10$^{3}$ (Harrison and Huang, 1990) and 10$^{4}$ yr (McLeod, 1996).
These values are somewhat higher than the order of magnitude time scale of 10$^{2}$
yr identified as the crossover in the sediment core and historical data.
The average power spectrum of the angular deviation from
the dipole from 25 simulations is shown in Figure 13. The spectrum has a
high-frequency region $S(f)\propto f^{-2}$ which slowly flattens out to
a constant spectrum at low frequencies. This is nearly consistent with the
spectra of inclination and declination from lake sediment time series shown
in Figures 8 and 9. The $S(f)\propto f^{-2}$ begins to flattens out at
a time scale roughly equivalent to the time scale of the intensity spectrum to
cross over from $1/f$ to $1/f^{2}$ behavior. The measured value of this crossover
in the lake sediment power spectra is 3 kyr. This value is consistent
with estimates of 10$^{3}$ to 10$^{4}$ years for the diffusion time across the core
from Harrison and Huang (1990) and McLeod (1996). A major discrepancy between
the model and the observed spectra is the absence of a flattening out of the
spectrum at high frequencies in the model calculation.  

The average spatial power spectrum 
of transects of $B_{z}$ for 25 simulations is given in Figure
14. The spectrum is $S(k)\propto k^{-1}$. This is consistent with the $R_{n}\propto n^{-1}$ power spectrum
from a spherical harmonic expansion as observed for the spatial power spectrum of the earth's 
magnetic field.  
The power spectrum of one-dimensional transects is directly comparable to the
power spectrum from a spherical harmonic expansion.  
For example, the power spectrum of the
earth's topography 
and bathymetry has been estimated from Fourier analysis of one-dimensional
transects and from 
a spherical harmonic expansion. The spectral exponents obtained are identical
in the two analyses (Turcotte, 1992). 
 
Although this model reproduces many of the observed features of the variability
of the earth's magnetic field, the physical origin of the terms in the
model equation have not been specified. Gaussian white noise amplification 
and decay of the magnetic field by dynamo action is a simple stochastic 
model for dynamo action but there is no physical justification for it. 
Moreover, we have presented no clear physical justification for 
the term $p-B_{z,tot}$ which generates field reversals and a binormal
distribution of dipole intensity. 
These terms were included in the model in order to construct a minimal model
consistent with the complex behavior of the earth's magnetic field. 

\section{Conclusions}
We have presented the results of power spectral analyses of variations in the
dipole moment of the earth's magnetic field which show the spectrum to be proportional
to $1/f$ from time scales of 100 yr to 4 Myr. We have also performed spectral analyses
of variations in inclination and declination. These spectra are  
constant above time scales
of 3 kyr, proportional to $1/f^{2}$ from time scales of 500 yr to 3 kyr,
and constant again below time scales of 500 yr. The 3 kyr time scale is
associated with the decay time of the quadrupole moment.
We have shown that reversals generated by 
binormal, $1/f$ noise geomagnetic field intensity variations
are consistent with the distribution of polarity lengths and clustering of
real reversals. We have shown how a model of magnetic diffusion driven by
dynamo action parameterized as a stochastic process reproduces many of the observed
features of the spatial and temporal variability of the earth's magnetic field. 

\acknowledgments

We wish to thank Donald L. Turcotte for helpful conversations
and Laure Meynadier for access to the marine sediment data and helpful
conversations regarding its interpretation.


\newpage

Barton, C.E., Spectral analysis of palaeomagnetic time series
and the geomagnetic spectrum, {\it Phil. Trans. R. Soc. London A}, {\bf 306},
203-209, 1982.

Blakely, R.J., Geomagnetic reversals and crustal spreading rates during the
Miocene, {\it J. Geophys. Res.}, {\bf 79}, 2979-2985, 1974.

Cande, S.C. and Kent, D.V., A new geomagnetic polarity time scale
for the Late Cretaceous and Cenozoic, {\it J. Geophys. Res.}, {\bf 97},
13,917-13,951, 1992a.

Cande, S.C. and Kent, D.V., Ultrahigh resolution marine magnetic anomaly
profiles
: a record of continuous paleointensity variations?, {\it J. Geophys. Res.},
{\bf 97}, 15075-15083, 1992b.

Cande, S.C. and Kent, D.V., Revised calibration of the geomagnetic polarity
timescale for the Late Cretaceous and Cenozoic, {\it J. Geophys. Res.}, {\bf 100},
6093-6095, 1995. 

Courtillot, V. and  Le Mouel, J.L., Time variations of the earth's magnetic
field: From daily to secular, {\it Ann. Rev. Earth Plan. Sci.}, {\bf 16},
389-476, 1988.
 
Cox, A., Lengths of geomagnetic polarity intervals, {\it J. Geophys. Res.},
{\bf 73}, 3247-3260, 1968. 

Currie, R.G., Geomagnetic spectrum of internal origin and lower
mantle conductivity, {\it J. Geophys. Res.}, {\bf 73}, 2779-2768, 1968.

Gaffin, S., Analysis of scaling in the geomagnetic polarity reversal record,
{\it Phys. Earth Plan. Inter.}, {\bf 57}, 284-290, 1989.

Harland, W.B., Cox, A., Llewellyn, P.G., Pickton, C.A.G., Smith, A.G., and 
Walters, R., 1990. {\it A Geologic Time Scale 1989},
Cambridge University Press,
London, 1989.

Harrison, C.G.A. and Huang, Q., Rates of change of the Earth's magnetic
field measured by recent analyses, {\it J. Geomag. Geoelectr.}, {\bf 42},
897-928, 1990.  

Kono, M., Intensity of the earth's magnetic field during the Pliocene and Pleistocene
in relation to the amplitude of mid-ocean ridge magnetic anomalies, {\it Earth and Plan.
Sci. Lett.}, {\bf 11}, 10-17, 1971.

Kono, M., Rikitake two-disk dynamo and paleomagnetism, {\it Geophys. Res. Lett.},
{\bf 14},
21-24, 1987.

Kovacheva, M., Summarized results of the archeomagnetic investigation of the
geomagnetic field variation for the last 8000 yr in south-eastern Europe,
{\it Geophys. J. R. Astr. Soc.}, {\bf 61}, 57-64, 1980.

Laj, C., Kissel, C., Lefevre, I., Relative geomagnetic field intensity 
and reversals from Upper Miocene sections in Crete, {\it Earth Plan. Sci.
Lett.}, {\bf 141}. 67-78, 1996.  

Langel, R.A. and Estes, R.H., 
A geomagnetic field spectrum, {\it Geophys. Res. Lett.}, {\bf 9},
250-253, 1982.

Lehman, B., Laj, C., Kissel, C., Mazaud, A., Paterne, M., and Labeyrie, L., 
Relative changes of the geomagnetic field intensity during the last 280 kyear
from piston cores in the Acores area, {\it Phys. Earth Plan. Int.}, {bf 93},
269-284, 1996.

Lock, J. and M.W. McElhinney, The Global Paleomagnetic Database: design,
installation, and use with ORACLE, {\it Surveys in Geophysics}, {\bf 12},
317-506, 1991.

Lund, S.P. and Banerjee, S.K., The paleomagnetic record of Late 
Quaternary secular variation from Anderson Pond, Tennessee, {\it Earth
Plan. Sci. Lett.}, {\bf 72}, 219-237, 1985.

Lund, S.P., Liddicoat, J.C., Lajoie, K.R., Henyey, T.L., and Robinson,
S.W., Paleomagnetic evidence for long-term (10$^{4}$ year) memory
and periodic behavior in the earth's core dynamo process, {\it Geophys.
Res. Lett.}, {\bf 15}, 1101-1104, 1988.    

McFadden, P.L. and Merrill, R.T., History of the Earth's magnetic field
and possible connections to core-mantle boundary processes, {\it J. Geophys.
Res.}, {\bf 100}, 307-316, 1995. 

McLeod, M.G., Signals and noise in magnetic observatory annual means:
Mantle conductivity and jerks, {\it J. Geophys. Res.}, {\bf 97}, 17,261-17,290, 1992.

McLeod, M.G., Spatial and temporal power spectra of the geomagnetic field,
{\it J. Geophys. Res.}, {\bf 101}, 2745-2763, 1996.

Meynadier, L., Valet, J.-P., Bassonot, F.C., Shackleton, N.J. and 
Guyodo, Y., Asymmetrical saw-tooth pattern of the geomagnetic field
intensity from equatorial sediments in the Pacific and Indian oceans,
{\it Earth Plan. Sci Lett.}, {\bf 126}, 109-127, 1994.

Meynadier, L., Valet, J.-P., Weeks, R., Shackleton, N.J. and Hagee, V.L., 
Relative geomagnetic intensity of the field during the last 140 ka, {\it
Earth Plan. Sci. Lett.}, {\bf 114}, 39-57, 1992.

Pal, P.C. and Roberts, P.H., Long-term polarity stability and strength of the
geomagnetic dipole, {\it Nature}, {\bf 331}, 702-705, 1988.

Pelletier, J.D. and Turcotte, D.L., Scale-invariant topography and porosity variations in
sedimentary basins, {\it J. Geophys. Res.}, {\bf 101}, 28,165-28,175, 1996. 

Press, W.H., Teukolsky, S.A., Vetterling, W.T. and Flannery B.P., {\it
Numerical Recipes in C: The Art of Scientific Computing}, second ed.,
Cambridge
University Press, Cambridge, 1992.

Seki, M. and Ito, K., A phase-transition model for geomagnetic polarity
reversals, {\it J. Geomag. Geoelectr.}, {\bf 45}, 79-88, 1993.

Stevenson, D.J., Planetary magnetic fields, {\it Rep. Prog. Phys.}, {\bf 46},
555-620, 1983.

Tauxe, L. and Hartl, P., 11 million years of Oligocene geomagnetic
field behavior, {\it Geophys. J. Int.}, {\bf 128}, 217-229, 1997.    

Thibal, J., Pozzi, J.-P., Barthes, V., and Dubuisson, G., Continuous record
of geomagnetic field intensity between 4.7 and 2.7 Ma from downhole measurements,
{\it Earth Plan. Sci. Lett.}, {\bf 136}, 541-550, 1995.

Thouveny, N., Creer, K.M., and Blunk, I., Extension of the Lac du
Bouchet palaeomagnetic record over the last 120,000 years, {\it Earth Plan.
Sci. Lett.}, {\bf 97}, 140-161, 1990.

Turcotte, D.L., {\it Fractals and Chaos in Geology and Geophysics},
Cambridge Univ. Press, Cambridge, 1992.

Turner, G.M. and Thompson, R., Lake sediment record of the geomagnetic 
secular variation in Britain during Holocene times, {\it Geophys. J. R.
Astron. Soc.}, {\bf 65}, 703-725, 1981. 

Valet, J.-P. and Meynadier, L., Geomagnetic field intensity and reversals
during the past four million years, {\it Nature}, {\bf 366}, 234-238, 1993.

van der Ziel, A., On the noise spectra of semiconductor noise and of flicker
effect, {\it Physica}, {\bf 16}, 359-375, 1950.

Verosub, K.L., Mehringer, P.J., and Waterstraat, P., Holocene secular 
variation in western North America: paleomagnetic record from Fish Lake,
Harney County, Oregon, {\it J. Geophys. Res.}, {\bf 91}, 3609-3623, 1986.

Vicsek, T., {\it Fractal Growth Phenomena}, World Sci., River Edge, N. J., 1992.

Voorhies, C.V. and Conrad, J., Accurate predictions of mean geomagnetic
dipole excursion and reversal frequencies, mean paleomagnetic field intensity,
and the radius of the Earth's core using McLeod's rule, {\it NASA Technical
Memorandum 104634}, 1996.

Weissman, M.B., $1/f$ noise and other slow, nonexponential kinetics in condensed
matter, {\it Rev. Mod. Phys.}, {\bf 60}, 537-571, 1988.

\newpage

\section*{Figure Captions}

Figure 1: Paleointensity of the virtual axial dipole moment (VADM) of the
earth's magnetic field (with reversed polarity data
given by negative values) 
inferred from sediment cores for the past 4 Ma from Meynadier (1994).

Figure 2: Power spectrum 
of the geomagnetic field intensity variations estimated with
the use of the Lomb periodogram from sediment
cores of Meynadier (1992) and Meynadier (1994) and archeomagnetic data from
Kovacheva (1980). The power spectrum $S$ is given as a function of  
frequency $f$ for time scales of 100 a to 4 Ma. 

Figure 3: Cumulative
frequency-length distribution of the lengths of polarity intervals during the
last 170 Ma from the
time scale of Harland et al. (1990) (solid curve), Cande and
Kent (1992a,1995) (dashed curve), and the Cande and Kent (1992a,1995)
time scale from C1 to C13 with cryptochrons included (dashed and dotted line).

Figure 4: (a) 
A $1/f$ noise with a normal distribution with mean of 8.9x10$^{22}$Am$^{2}$
and standard
deviation of 3.4x10$^{22}$Am$^{2}$ representing the geomagnetic field intensity (VADM) in
one polarity state.
(b) Binormal $1/f$ noise constructed from the normal $1/f$ noise of (a)
as described in the text.

Figure 5: Cumulative frequency-length polarity interval distributions from the
Harland et al. (1990) time scale and that of the binormal, $1/f$ noise model of
intensity variations. The distribution from the Harland et al. (1990) time scale
(dashed curve) was also given in Figure 4. The solid line represents the
average cumulative distribution from the $1/f$ noise model. The dotted lines
represent the minimum and maximum reversal length distributions for 20 numerical
experiments, thereby representing 95\% confidence intervals. 

Figure 6: Cumulative frequency-length polarity interval distributions for the
$1/f$ noise model of intensity variations (shown in the middle, also given
in Figure 5) and for intensity variations with power spectra proportional
to $f^{-0.8}$ and $f^{-1.2}$. This plot
illustrates that the polarity length distribution is very sensitive to the form
of the power spectrum, allowing us to conclude that the agreement between the
model and the observed distribution in Figure 5 is unique to $1/f$ noise intensity
variations.

Figure 7: Pair correlation function of the reversal history according to the
Harland et al. (1990) time scale (filled circles), Cande and Kent (1992a,1995) (unfilled
circles), synthetic reversals produced from $1/f$ noise 
model of intensity variations (boxes), and a Poisson process (triangles). 
The real and synthetic reversals exhibit similar non-random clustering.

Figure 8: Power spectra of inclination and declination from the Lac du
Bouchet sediment core. The declination spectrum is offset from the inclination
spectrum so that they may be placed on the same graph.
 
Figure 9: Power spectra of inclination from the following locations,
top to bottom: 1) Anderson Pond, 2) Bessette Creek, 3) Fish Lake, 4)
Lake Bullenmerri, and 5) Lake Keilambete. The spectra are offset to place
them on the same graph.

Figure 10: Dipole moment produced by the model  
normalized to the average dipole moment, set to be one. 
The field exhibits reversals
with a broad distribution of polarity interval lengths and a variable reversal
rate decreasing at later times in the simulation.
 
Figure 11: Distribution of magnetic field according to the ten simulations of
the model (solid curve)
and a binormal distribution fit to the data (dashed line).
The binormal distribution fits
the data well.
 
Figure 12: Average power spectrum of the mean value of the magnetic field
(dipole field) from 25 simulations. The spectrum has a low-frequency spectrum
with $S(f)\propto f^{-1}$ and a high frequency region $S(f)\propto f^{-2}$.
The same spectrum is observed in geomagnetic intensity from sediment cores
and historical data.
 
Figure 13: Average power spectrum of the angular deviation from a dipole
field from 25 simulations. The spectrum is $S(f)\propto f^{-2}$ for high
frequencies and gradually flattens out to a constant spectrum at low
frequencies.

Figure 14: Average spatial power spectrum of transects of the magnetic field 
$B_{z}$. The power spectrum is $S(k)\propto k^{-1}$. This is precisely analagous
to the power-law spectrum $R_{n}\propto n^{-1}$.  

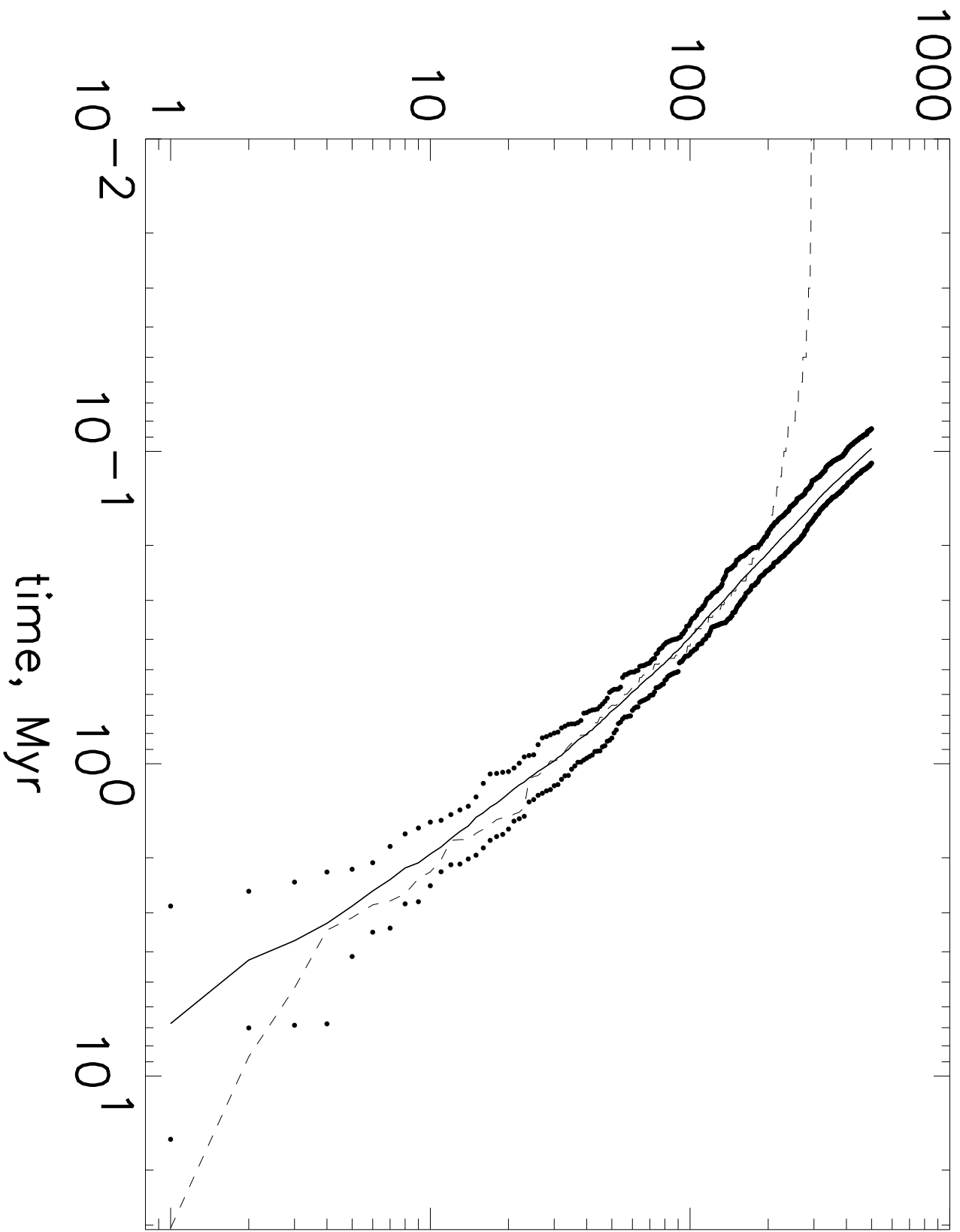
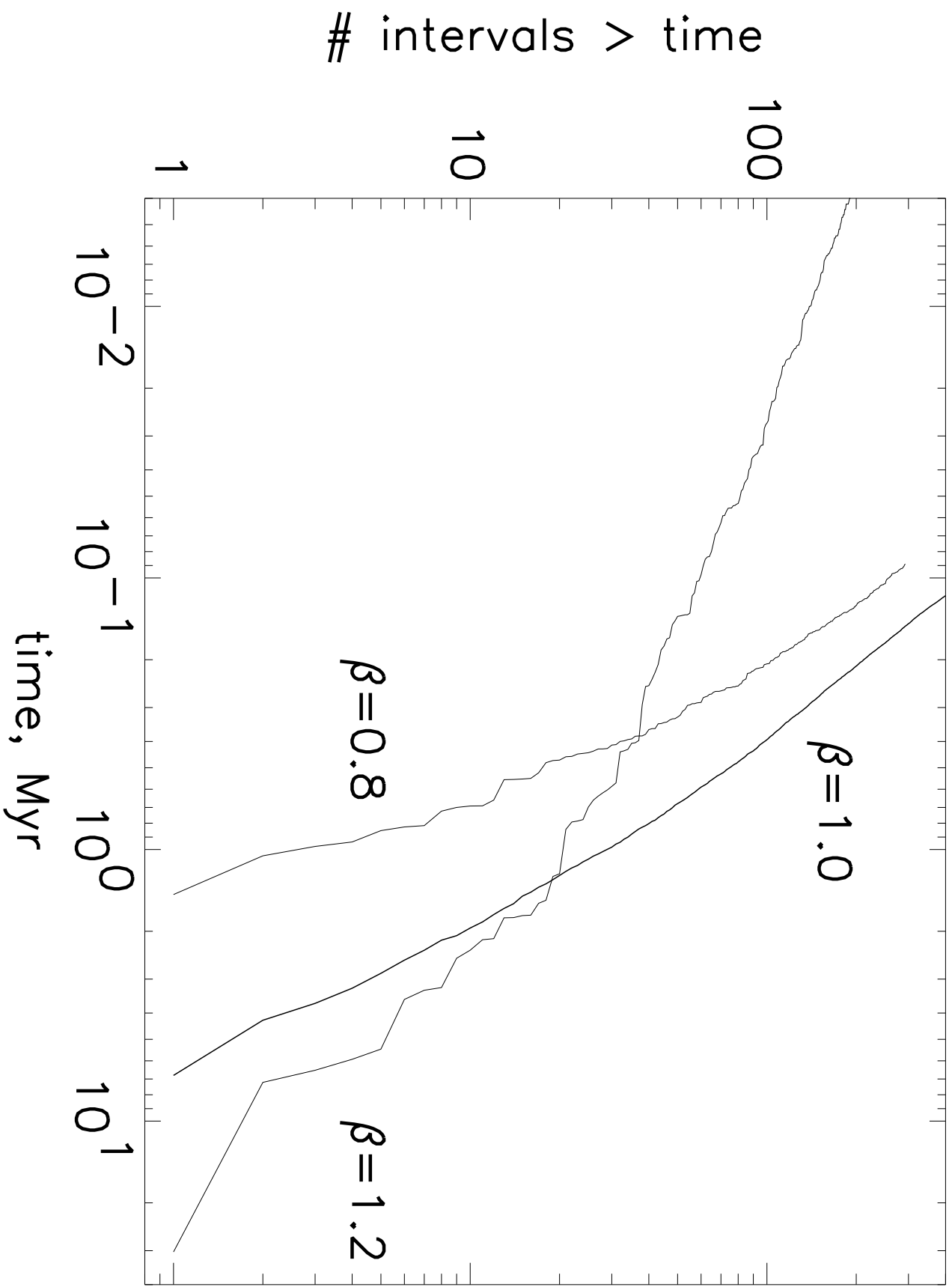
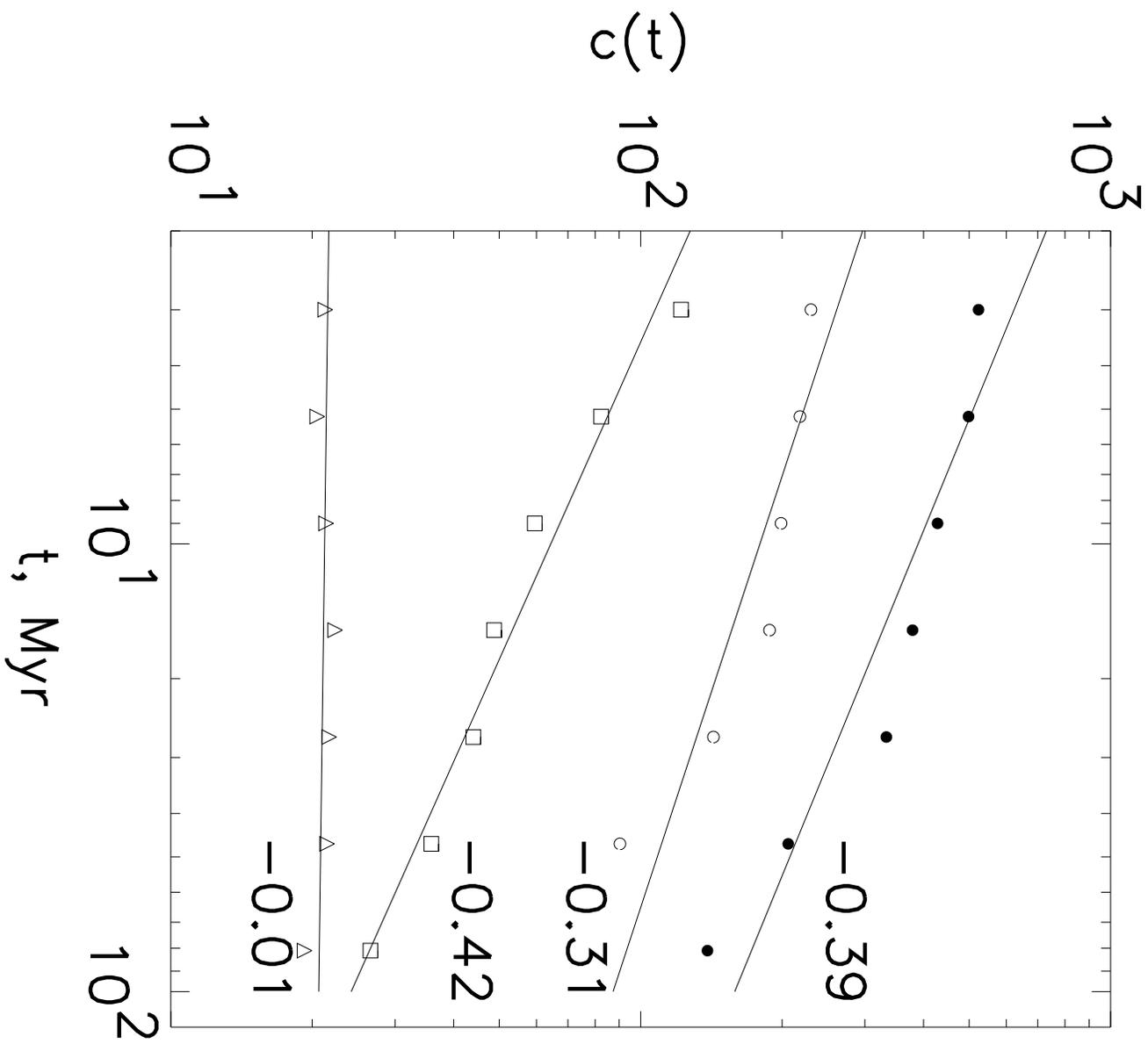
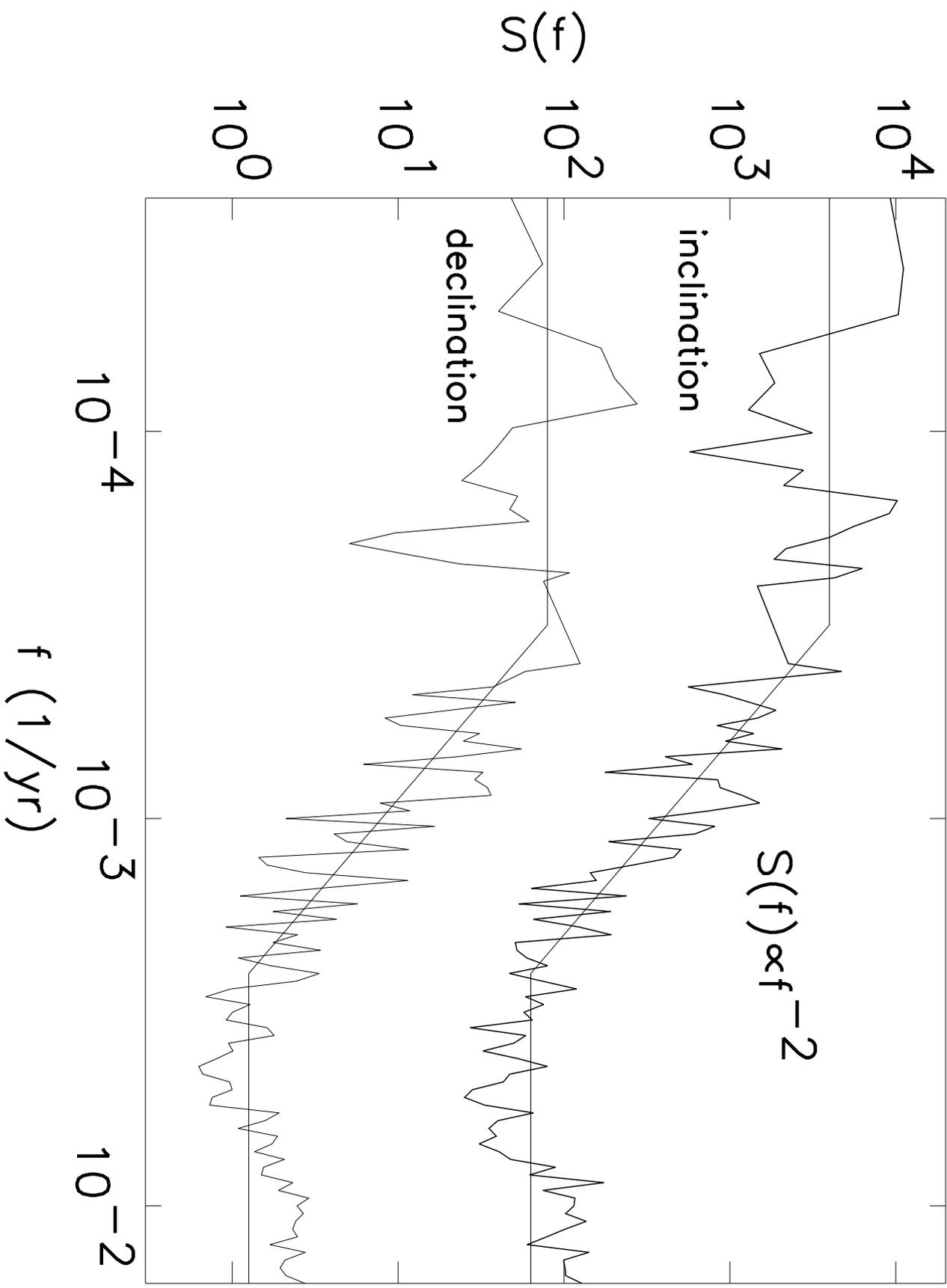
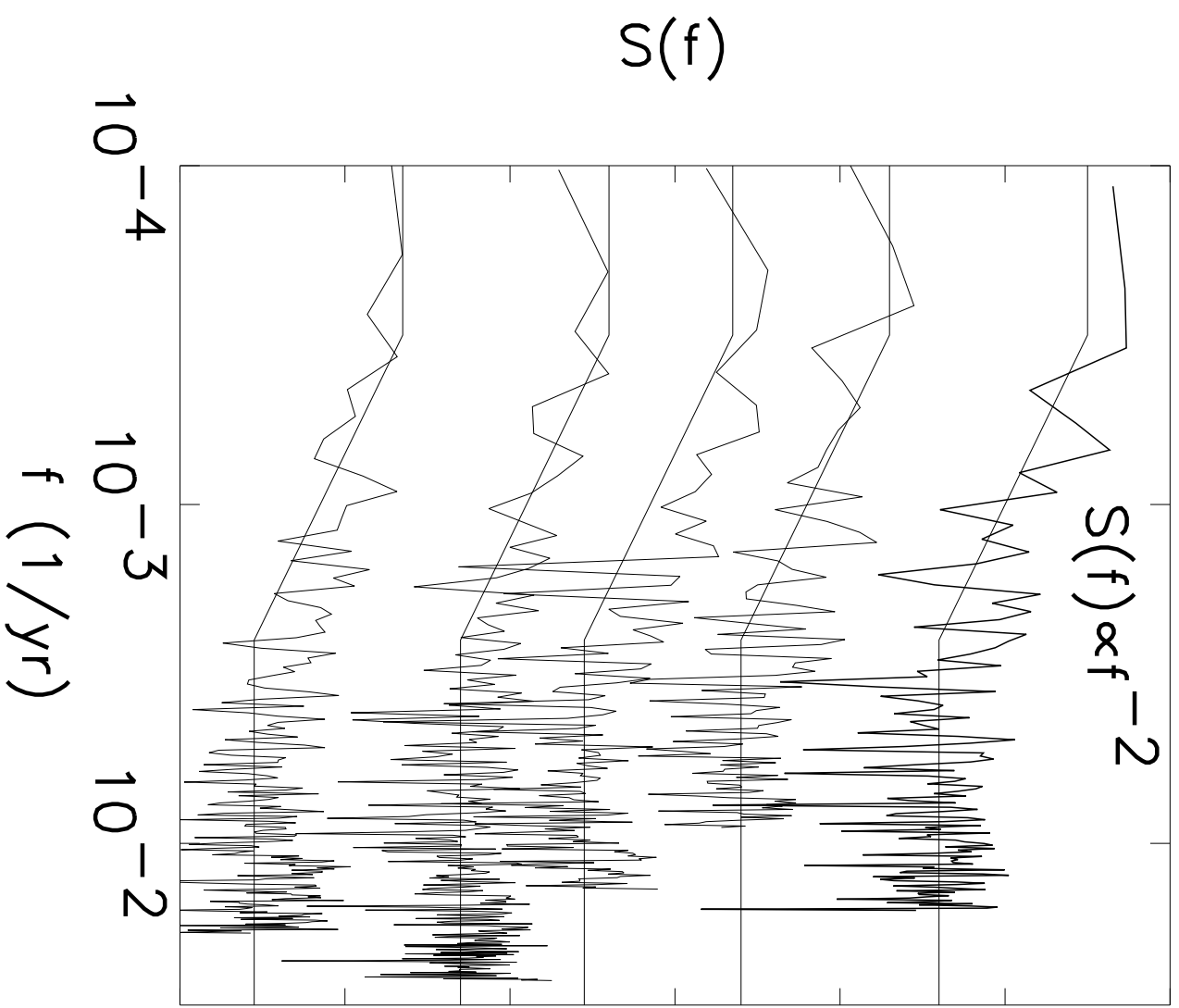
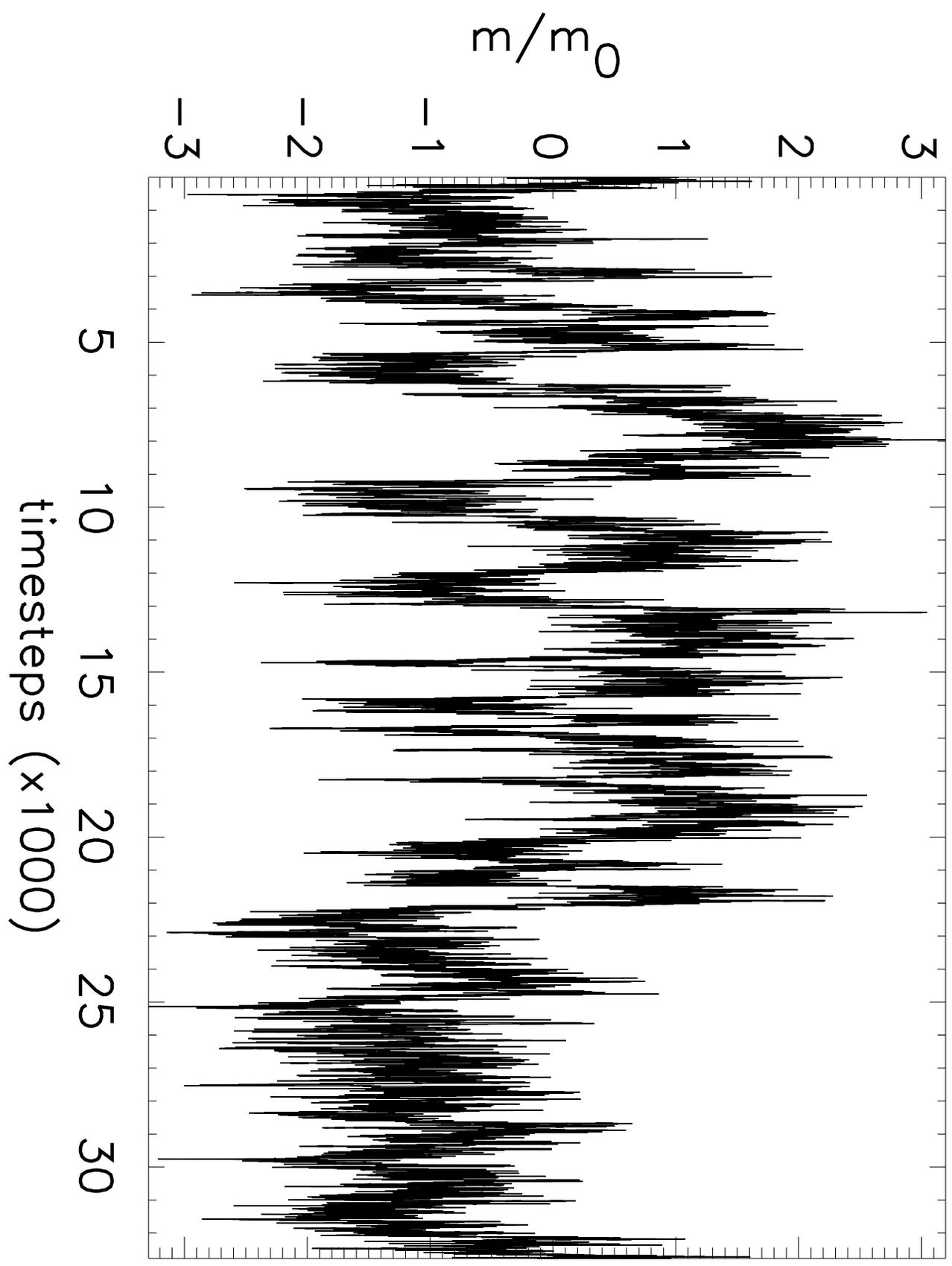
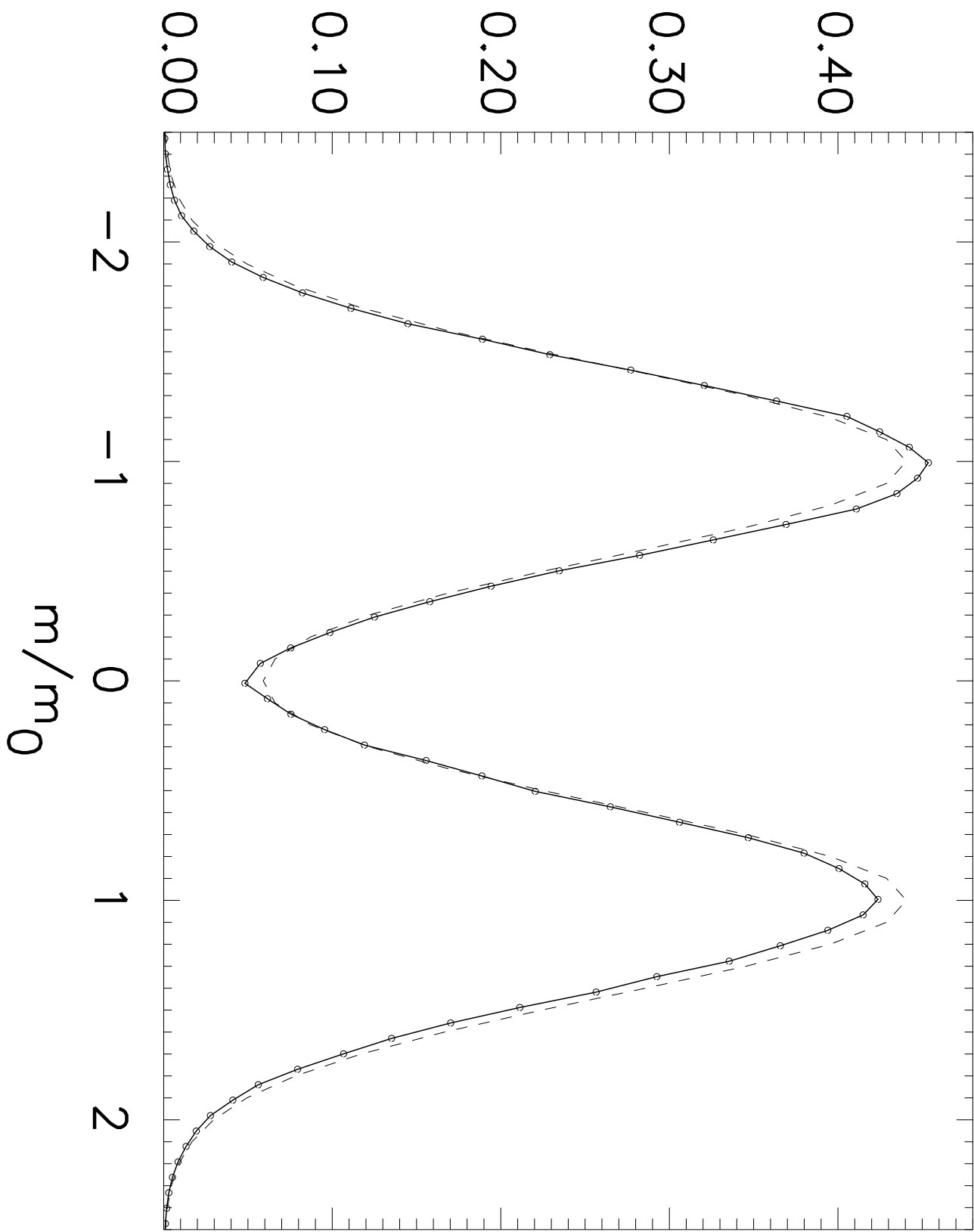
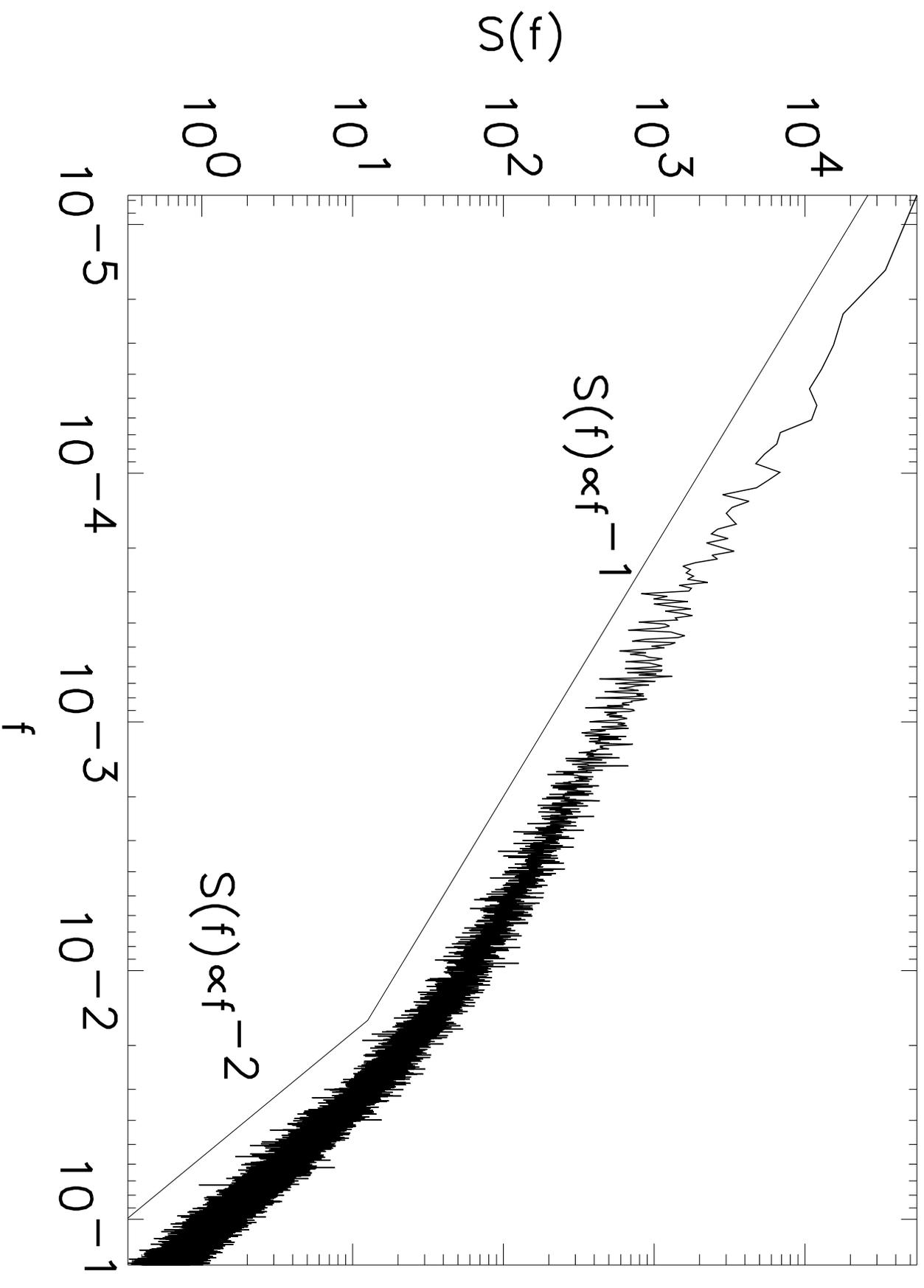
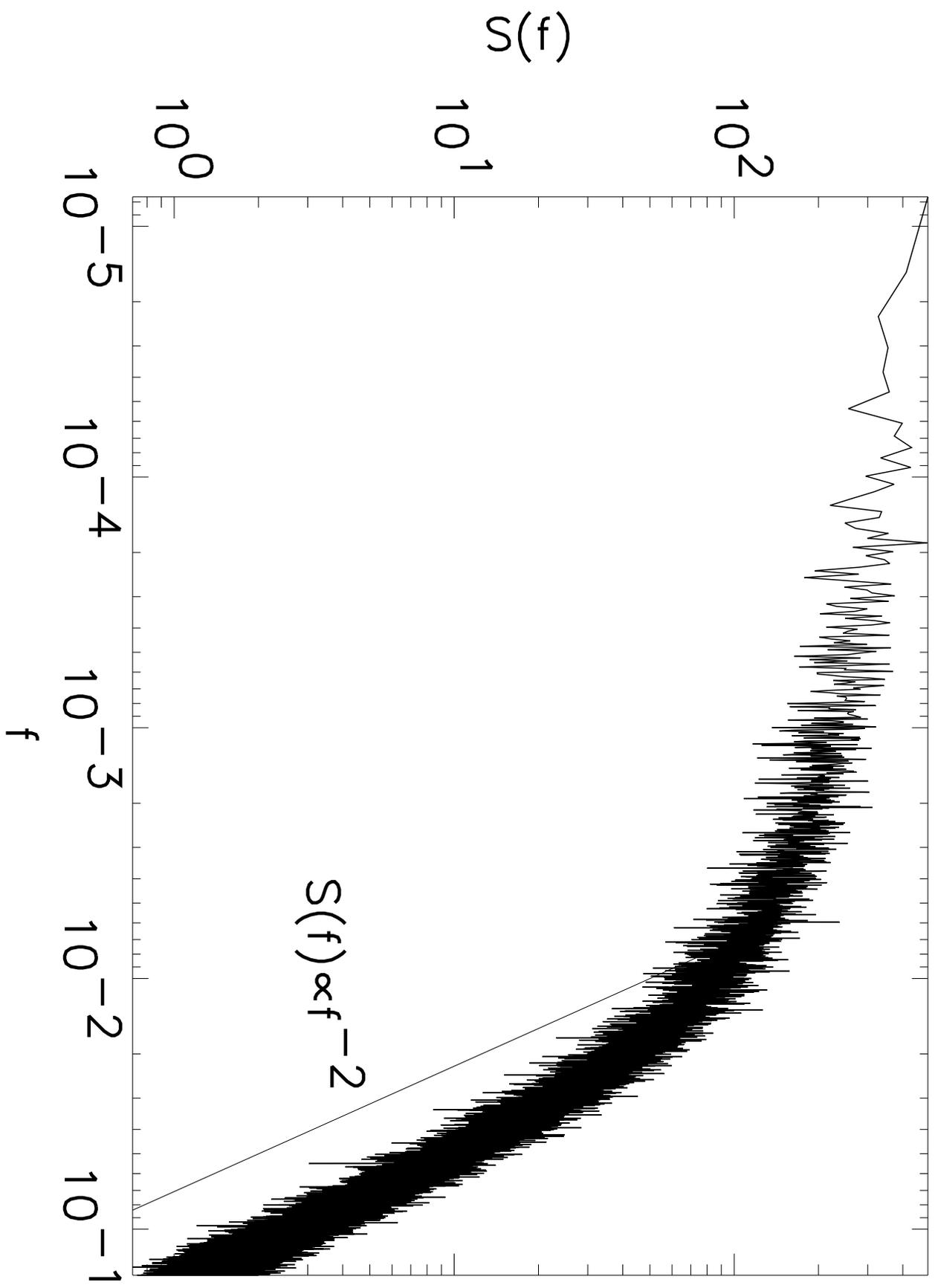
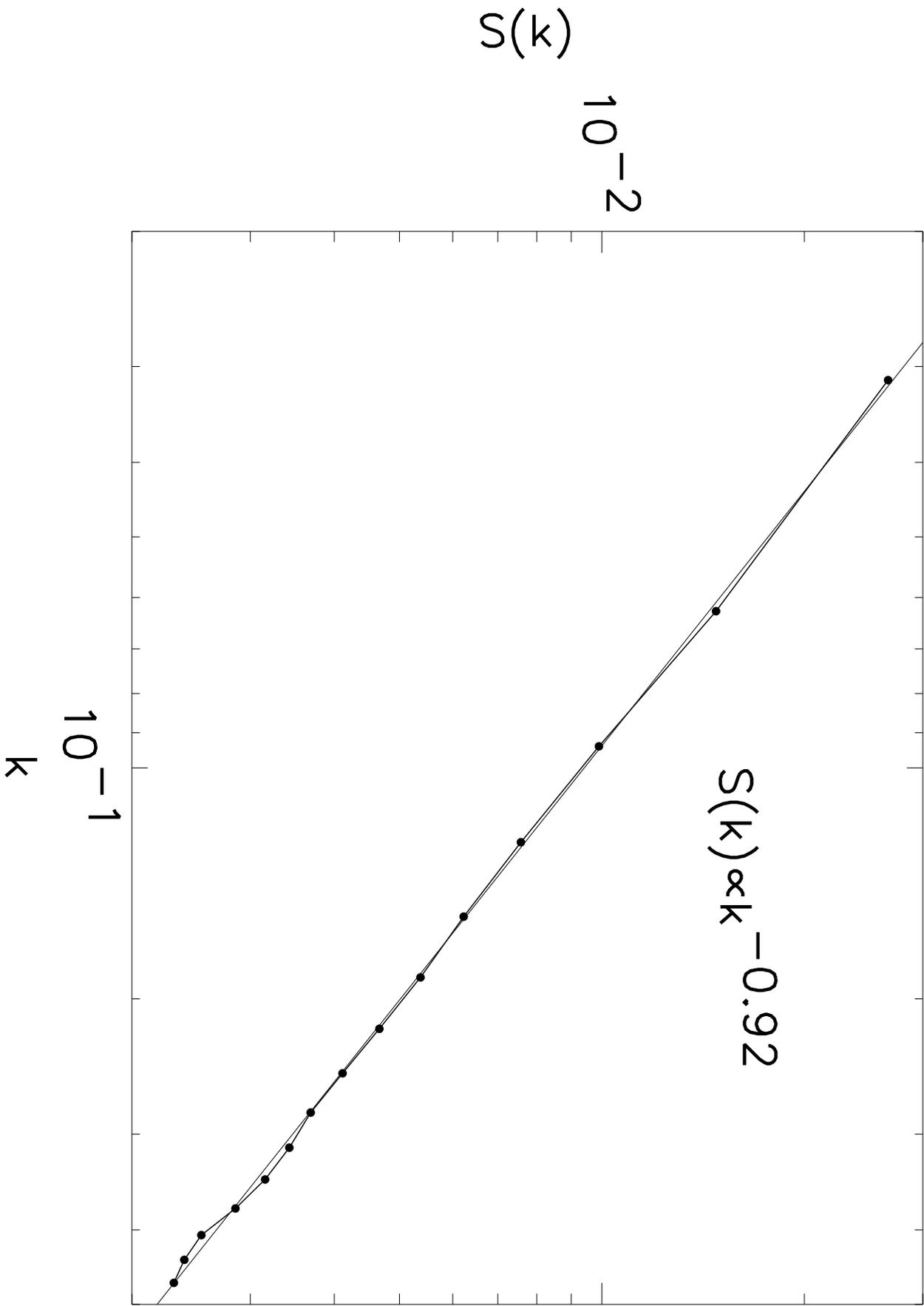
\end{document}